\documentstyle[aas2pp4,iau200p]{article}

\newcommand{\aas} [2]{A\&A Suppl. #1, #2}

\begin{document}
\title{Stringent constraints on single-star  
evolution theory from masses and radii of well-detached 
binaries in star clusters}
\author{Erwan Lastennet}
\affil{Astronomy Unit, Queen Mary and Westfield College, \\
       Mile End Road, London E1 4NS, UK \\
       {\tt E.Lastennet@qmw.ac.uk}}
\and
\author{David Valls-Gabaud}
\affil{Laboratoire d'Astrophysique, UMR CNRS 5572, \\
       Observatoire Midi-Pyr\'en\'ees, 14 avenue Belin, \\
       F-31400 Toulouse, France \\
       {\tt dvg@ast.obs-mip.fr}}
\and
\author{Edouard Oblak}
\affil{CNRS UPRESA 6091, Observatoire de Besan\c con, \\
       41 bis avenue de l'Observatoire, \\
       F-25010 Besan\c con, Cedex, France \\
       {\tt oblak@obs-besancon.fr}}

\begin{abstract}
Which sample of objects can give strong constraints on single-star 
evolution theory ? Whilst star cluster members share the same age 
and the same metallicity, many questions (e.g. field stars contamination, 
stellar rotation, presence of unresolved binaries) are difficult to 
clarify properly when using their colour-magnitude diagram to compare 
with theoretical isochrones. 
Alternatively, binary stars can be used to put constraints on theoretical 
predictions. However, while the stellar mass is accurately known for 
some sample of well-detached binary stars, their metallicities are often 
poorly known. 
Then it appears that a better test could be obtained by combining both 
advantages (well-detached binaries members of a star cluster). This 
idea is applied in this work to binaries in the Hyades and one binary  
in the Cepheus OB3 association to test the validity of three independent 
sets of theoretical tracks. A detailed comparison of theoretical vs. 
observational masses (and radii when possible) are presented. 

\end{abstract}

\section{Introduction}

Binary systems are the main source of fundamental data on stellar masses and radii. 
These data give stringent constraints that should be fitted by any set of 
theoretical stellar models. 
This work is exclusively focused on the mass estimate (and radius when available) of 
well-detached binary systems which can be presumed typical of single stars 
properties. Moreover three of them should share same age, chemical composition and 
distance due to their membership to the same open cluster. 
The 6 selected Hyades stars have individual stellar masses known with an accuracy of 
about 10\% for 51 Tauri and $\theta^2$ Tauri and better than 1\% for V818 Tauri. 
The mass accuracy is better than 3\% for the CW Cephei eclipsing binary (Table~\ref{tab:par}). 
Moreover, these stars cover a wide mass range which is useful to obtain some interesting 
tests between $\sim$0.77 and 13.5$M_{\odot}$. 

\begin{table}[htb] 
  \caption[]{Cross identification and masses for the 4 selected binary systems.}  
  \label{tab:par}
  \smallskip
  \begin{tabular}{lrcl}
  \hline \hline
  Stars            &  HIP   & Mass            & Ref. \\
                   &        & ($M_{\odot}$)   &      \\
  \tableline
  51 Tau A         &  20087 &  1.80 $\pm$ 0.13     & [TSL97a] \\ 
  51 Tau B         &  ...   &  1.46 $\pm$ 0.18     & [TSL97a] \\
  V818 Tau A       &  20019 &  1.072 $\pm$ 0.010   & [PS88] \\
  V818 Tau B       &  ...   &  0.769 $\pm$ 0.005   & [PS88] \\
  $\theta^2$ Tau A &  20894 &  2.42 $\pm$ 0.30     & [TSL97b]$^{(1)}$ \\
  $\theta^2$ Tau B &  ...   &  2.11 $\pm$ 0.17     & [TSL97b]$^{(2)}$ \\
  CW Cep A         & 113907 &  13.52 $\pm$ 0.39    & [A91] \\
  CW Cep B         &  ...   &  12.08 $\pm$ 0.29    & [A91] \\
  \hline \hline
  \end{tabular}
  
  $^{(1)}$ The value originally quoted by [TSL97a], 2.10$\pm$0.60 $M_{\odot}$, 
  is the determination of Tomkin et al. (1995) adjusting the error upward by a 
  factor of two. \\
  $^{(2)}$ The value quoted by [TSL97a], 1.60$\pm$0.40 $M_{\odot}$, 
  is also from Tomkin et al. (1995) adjusting the error upward by a factor of two.\\
\end{table}

Comparisons between measured masses and predictions of three widely used models 
from the Geneva group (Mowlavi et al. 1998 and references therein), the Padova group 
(Fagotto et al. 1994 and references therein) and from Claret \& Gim\'enez (1992) 
(CG92 thereafter) are presented in the next sections. 
The theoretical quantities are derived from the isochrone technique in the colour 
magnitude diagram as described in Lastennet et al. (1999). 

\section{HIP 20087 $=$ 51 Tauri}

The 51 Tau system is a spectroscopic binary and also a visual binary resolved by 
speckle interferometry. 
Since the masses of the two components of 51 Tau are known to a good accuracy 
(respectively 7\% and 12\%), it is essential to check that the predicted masses are in 
agreement with the measured masses.  
The best Geneva fit slightly underestimates the masses of 51 Tau (by 0.5$\sigma$ 
for the primary and 1$\sigma$ for the secondary component) as do the Padova fit
(respectively 1$\sigma$ for the primary and $\sim$1.2$\sigma$ for the secondary) 
and CG92 (respectively 0.5$\sigma$ for the primary and 1$\sigma$ for the secondary).  
This means that these models are able to reproduce to a satisfying level the stellar 
masses in the mass range of 51 Tau (see Table~\ref{tab:par1} for details). 

\begin{table}[htb] 
  \caption[]{Comparison of theoretical mass estimates from isochrone fitting 
             of the system 51 Tau with the values of Torres et al. [TSL97a].}
  \label{tab:par1}
  \smallskip
  \begin{tabular}{ccccc}
  \hline \hline
  & [TSL97a]  & Geneva$^{a}$ & Padova$^{a}$ & CG92$^{a}$ \\
  \tableline 
  A & 1.80$\pm$0.13 & 1.74$\pm$0.06 & 1.67$\pm$0.04 & 1.72$\pm$0.05 \\
  B & 1.46$\pm$0.18 & 1.27$\pm$0.05 & 1.24$\pm$0.04 & 1.28$\pm$0.05 \\
  \hline \hline
  \end{tabular}
$^{a}$ this work. \\
\end{table}

The agreement is even better if we take into account 
the following point. When spectroscopic and astrometric orbital elements are known, 
one can derive the mass of each component adopting a distance (i.e. the parallax). 
Hence, if we keep all the orbital parameters (period, eccentricity, 
velocity amplitude, ...) as given in [TSL97a] and take the Hipparcos parallax value of 
51 Tau (ESA, 1997), one derives (from Eq. 2 in [TSL97b]) the new masses 
M$_A$$\simeq$1.66$M_{\odot}$ and M$_B$$\simeq$1.40$M_{\odot}$, values that are even 
better matched by the three sets of theoretical models.

\section{HIP 20019 $=$ V818 Tauri}

The V818 Tau system is a double-lined eclipsing binary (Mc Clure, 1982) with very 
well estimated masses (actually the most accurate masses known for Hyades members). 
Indeed, the relative errors on the masses are less than 1\% 
(Peterson \& Solensky, 1988, [PS88]), and the secondary component 
is particularly interesting because it is one of the rare stars which is less massive than 
the Sun and whose mass is known with such high accuracy (cf. Andersen 1991 for a comprehensive 
review of accurate data in double-lined eclipsing binaries). Unfortunately, 
this low mass star (about 0.77 $M_{\odot}$) does not allow us to test either the CG92 models 
-- the lower available mass of these models being 1 $M_{\odot}$ -- or the Geneva 
models whose lower mass limit is 0.8 $M_{\odot}$.

Since the masses of the two components of V818 Tau are known with an excellent accuracy, 
checking that the stellar masses predicted by models are in agreement with the true 
masses is a critical test. A comparison is given in Table~\ref{tab:par2}. 
The result is that the best Padova fit underestimates 
the masses of V818 Tau by about 5$\sigma$ for both stars. This is not so bad because 
of the high level of accuracy of the masses: 5$\sigma$ only means 0.05$M_{\odot}$ 
for V818 Tau A and 0.025$M_{\odot}$ for V818 Tau B. \\

\begin{table}[htb] 
  \caption[]{Comparison of theoretical mass estimates from Padova isochrone fitting 
  of the system V818 Tau with the values of [SM87] (Schiller \& Milone, 1987) and [PS88] 
  (Peterson \& Solensky, 1988).}
  \label{tab:par2}
  \smallskip 
  \begin{tabular}{cccc}
  \hline \hline
  & [SM87] & [PS88]  & Padova$^{a}$ \\  
  \tableline
  A & 1.080$\pm$0.017 & 1.072$\pm$0.010 & 1.020$\pm$0.04  \\
  B & 0.771$\pm$0.011 & 0.769$\pm$0.005 & 0.744$\pm$0.02  \\
  \hline \hline
  \end{tabular}
  
$^{a}$ this work. \\
\end{table}

\begin{figure}[t]
\plotfiddle{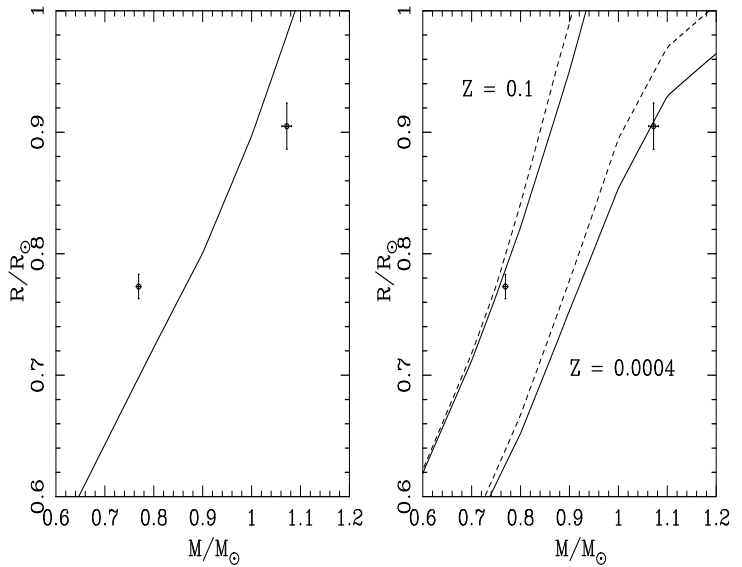}{8.5 cm}{-90}{35}{45}{-150}{240}
\caption{
V818 Tau system in a Mass-Radius diagram where all the isochrones are derived from 
the Padova tracks. 
The mass of each star is from [PS88], as quoted by Torres et al. [TSL97a], and the radius is 
from [SM87]. 
In the {\it left panel}, the {\it solid line} isochrone (log t $\simeq$ 7.30, Z $\simeq$ 0.033) 
corresponds to the best fit isochrone in the (M$_V$, B$-$V) colour magnitude diagram (see 
Lastennet et al. 1999 for all details). 
This isochrone clearly does not fit V818 Tau in the mass-radius diagram. 
In the {\it right panel}, four different isochrones are shown for comparison: with a high 
metallicity (Z $=$ 0.1) for two ages, t $=$ 0 ({\it solid line}) and t $=$ $10^9$ yrs ({\it dashed 
line}) and with a very low metallicity (Z $=$ 0.0004) for the ages t $=$ 0 ({\it solid line}) and 
t $=$ $10^9$ yrs ({\it dashed line}). There is no Padova isochrone which is able to fit 
simultaneously both components of V818 Tau in the mass-radius diagram, whatever the combination 
of age and metallicity.}
\label{fig1}
\end{figure}

Moreover, since V818 Tau is a double-lined eclipsing binary, the radius of each component 
is also known with an excellent accuracy (better than 2\%): 
R$_A$$=$ 0.905$\pm$0.019 $R_{\odot}$ and R$_B$$=$ 0.773$\pm$0.010 $R_{\odot}$ according to 
Schiller \& Milone (1987). On the other hand, the stellar radius can also be 
computed from the effective temperature and the luminosity given by the Padova models. 
Therefore, it is possible to compare the stellar radii predicted by the models with the true 
radii. 
Unfortunately, even if Padova models predict a fairly correct mass for each component 
of V818 Tau, it appears clearly from Fig. 1 that there is no Padova isochrone able 
to fit both components of V818 Tau in the mass-radius diagram. 
With the best fit isochrone (which is {\it not} a good fit, see Fig. 1), 
the radius of the more massive component is overestimated by $\sim$4$\sigma$, and the 
radius of the less massive component is underestimated by more than 7$\sigma$.   
This shows that tests using masses without taking radii into account are much less 
discriminant, hence the importance of double-lined eclipsing binaries to fully 
constrain stellar tracks.

\section{HIP 20894 $=$ $\theta^2$ Tauri}

The primary component of the spectroscopic binary $\theta^2$ Tau (spectral type A7 III) is 
one of the brightest stars in the Hyades. 
Such a system is {\it a priori} very critical because it is composed by a main-sequence star 
and an  evolved star, which allows us to test widely different evolutionary stages.
Since the masses of the two components of $\theta^2$ Tau are known to a good accuracy 
(respectively 12\% and 8\%), to check that the predicted masses are in 
agreement with the measured masses is a powerful test.

\begin{table}[htb] 
  \caption[]{Comparison of theoretical mass estimates from Geneva isochrone fitting 
  of the system $\theta^2$ Tau with the values of [TPM95] (Tomkin et al., 1995) 
  and [TSL97b].}
  \label{tab:par3}
  \smallskip 
  \begin{tabular}{cccc}
  \hline \hline
  & [TPM95] & [TSL97b]  & Geneva$^{a}$ \\ 
  \tableline 
  A & 2.1$\pm$0.3 & 2.42$\pm$0.30 & 2.37$\pm$0.02  \\
  B & 1.6$\pm$0.2 & 2.11$\pm$0.17 & 1.95$^{+0.06}_{-0.03}$  \\
  \hline \hline
  \end{tabular}
  
  $^{a}$ this work. \\
\end{table}

Torres et al. [TSL97b] updated the previous work on this system (Tomkin et al. 1995), 
which resulted in an increase of the masses (M$_A$ $=$ 2.42$M_{\odot}$ instead of 
M$_A$ $=$ 2.10$M_{\odot}$ and M$_B$ $=$ 2.11$M_{\odot}$ instead of 
M$_B$ $=$ 1.60$M_{\odot}$).
As shown in Table~\ref{tab:par3}, the best Geneva fit gives a quite 
good agreement  which means that these models are able to reproduce to a very satisfying 
level the stellar masses in the mass range of $\theta^2$ Tau. \\

\section{HIP 113907 $=$ CW Cephei} 

This massive system is a member of the Cepheus OB3 association (Clausen \& Gim\'enez 1991). 
Table~\ref{tab:par4} shows a very good agreement between the masses predicted and the true 
masses (Andersen 1991, [A91]).    

\begin{table}[htb] 
  \caption[]{Comparison of theoretical mass estimates from isochrone fitting 
             of the system CW Cep with the values of Andersen (1991).}
  \label{tab:par4}
  \smallskip
  \begin{tabular}{ccccc}
  \hline \hline
  & [A91]  & Geneva$^{a}$ & Padova$^{a}$ & CG92$^{a}$ \\
  \tableline 
  A & 13.52$\pm$0.39  & 13.48$^{+0.57}_{-0.54}$ & 13.52$^{+0.54}_{-0.53}$ & 13.44$^{+0.55}_{-0.53}$ \\
  B & 12.08$\pm$0.29  & 12.08$^{+0.56}_{-0.51}$ & 12.08$^{+0.54}_{-0.50}$ & 12.08$^{+0.54}_{-0.55}$ \\
  \hline \hline
  \end{tabular}
$^{a}$ this work. \\
\end{table}

While the Padova models fail to fit the radius of the low mass stars of V818 Tau 
at the level of accuracy required, they do predict correctly the individual radii 
of CW Cep: R$_A$$=$5.668$^{+0.219}_{-0.208}$$R_{\odot}$ and 
R$_B$$=$5.168$^{+0.263}_{-0.239}$$R_{\odot}$ 
are in agreement with the values listed in Andersen (1991),  
R$_A$$=$5.685$\pm$0.130$R_{\odot}$, R$_B$$=$5.177$\pm$0.129$R_{\odot}$.

\section{Conclusions}

The main conclusions of this work are the following: \\
1/ The masses predicted by the three sets of theoretical tracks, 
widely used in the literature, are in good agreement with the measured 
individual masses of each system in HIP 20087, HIP 20894 and HIP 113907. \\
2/ Padova isochrones can not fit simultaneously the two stellar components of 
V818 Tau in the mass-radius diagram, whatever the age and metallicity assumed. 
The radius of the more massive component is overestimated by $\sim$4$\sigma$, 
and the radius of the less massive component is underestimated by more than 7$\sigma$. \\
2a/ No conclusion on this point for the two other models because the low mass star 
($\sim$ 0.77 $M_{\odot}$) of V818 Tau does not allow us to test either the CG92 
or the Geneva models whose lower mass limit is 0.8 $M_{\odot}$. \\
3/ The result 2/ points out that the Padova models have to be used cautiously 
for accurate studies in the mass range $\sim$0.7-1.1 $M_{\odot}$.  \\
4/ It is of interest to keep in mind that mass and radius can be compared with 
theoretical predictions only for V818 Tau and CW Cep, the only double-lined eclipsing binaries of 
the sample studied in this work. 
If Padova tracks predict the true stellar masses of V818 Tau (but only at a 5$\sigma$ 
level), they are not able to reproduce simultaneously its true masses and radii with a 
great accuracy. Unfortunately, this result cannot be checked with the two other Hyades 
binaries, hence the importance of double-lined eclipsing binaries to fully constrain 
stellar tracks. \\
In this context, accurate data of well-detached double lined eclipsing binaries 
(e.g. Kurpinska-Winiarska et al. 2000) are highly needed to perform further detailed 
comparisons.  
Since 1996, Oblak and collaborators (see Oblak et al. 1999) started an observational 
campaign of radial velocity curves for a sample of new eclipsing binaries discovered 
by Hipparcos, and these new data will be of great interest for a better understanding 
of stellar evolution.

\end{document}